\documentclass[runningheads]{llncs}
\usepackage{bbding}

\usepackage{graphicx}
\usepackage{multirow}
\usepackage{booktabs}
\usepackage{diagbox}
\usepackage{bbm}
\usepackage{amsmath}
\usepackage{amssymb} 

\begin{document}

\title{Enable the Right to be Forgotten with Federated Client Unlearning in Medical Imaging}

%
%
\author{Zhipeng Deng\inst{1} \and
Luyang Luo\inst{1}\and
Hao Chen\inst{1,2,3}\Envelope }

\authorrunning{Z. Deng et al.}

\institute{Department of Computer Science and Engineering, \and
Department of Chemical and Biological Engineering, \and
Division of Life Science, \\
The Hong Kong University of Science and Technology, Hong Kong, China
\\
\email{zdengaj@connect.ust.hk}, \email{cseluyang@ust.hk}, \email{jhc@cse.ust.hk}}

\titlerunning{Federated Client Unlearning} 

\maketitle              

\begin{abstract}

The \textit{right to be forgotten}, as stated in most data regulations, poses an underexplored challenge in federated learning (FL), leading to the development of federated unlearning (FU). However, current FU approaches often face trade-offs between efficiency, model performance, forgetting efficacy, and privacy preservation. In this paper, we delve into the paradigm of Federated Client Unlearning (FCU) to guarantee a client the right to erase the contribution or the influence, introducing the first FU framework in medical imaging. In the unlearning process of a client, the proposed model-contrastive unlearning marks a pioneering step towards feature-level unlearning, and frequency-guided memory preservation ensures 
smooth forgetting of local knowledge while maintaining the generalizability of the trained global model, thus avoiding performance compromises and guaranteeing rapid post-training. We evaluated our FCU framework on two public medical image datasets, including Intracranial hemorrhage diagnosis and skin lesion diagnosis, demonstrating that our framework outperformed other state-of-the-art FU frameworks, with an expected speed-up of 10-15 times compared with retraining from scratch. The code and the organized datasets can be found at: https://github.com/dzp2095/FCU.
 

\end{abstract}
\section{Introduction}
To address the strict requirements on the collection, storage, and processing of personal data proposed in regulations like the General Data Protection Regulation (GDPR) \cite{voigt2017eu} and the California Consumer Privacy Act (CCPA) \cite{harding2019understanding}, federated learning (FL) \cite{mcmahan2017communication,Deng2023,liu2021federated,li2021fedbn} is regarded a promising privacy-preserving approach in medical imaging, which enables multiple parties to train a model collaboratively without sharing patient data. However, despite its decentralized nature, current FL research in medical imaging has not fully addressed the right to remove the influence of data from a trained global FL model, a right explicitly stated in GDPR as the \textit{right to be forgotten} and in CCPA as the \textit{right to delete}. In centralized learning, the right to have data removed can be realized by Machine Unlearning (MU) \cite{xu2023machine,cao2015towards}. Regardless, the existing MU techniques are developed for the centralized scenarios, posing significant challenges to their direct application in distributed FL settings \cite{halimi2022federated}, highlighting the need for dedicated federated unlearning (FU) approaches.

Retraining from scratch without the target forgotten data is regarded as a naive way to achieve unlearning \cite{bourtoule2021machine}. Nevertheless, this approach demands a large cost in communication and computation, especially in FL \cite{romandini2024federated}. Hence, recent studies have proposed various federated unlearning (FU) methods including re-calibration of historical updates \cite{liu2021federaser}, gradient quantization \cite{che2023fast}, gradient modification \cite{halimi2022federated} or knowledge distillation \cite{wu2022federated,zhao2023federated}. However, these methods often compromise performance or privacy. For instance, FedEraser\cite{liu2021federaser} accelerates retraining progress and removes the contribution of a target client iteratively by utilizing historical parameter updates of clients stored on the server side, yet this approach requires additional storage and poses a risk of data reconstruction by a malicious server \cite{singhal2021federated}.  FFMU \cite{che2023fast} applies randomized gradient smoothing and quantization to execute unlearning operations on the target forgotten data, but may fail to retain the performance when a client decides to remove all their data. UPGA \cite{halimi2022federated} formulates the unlearning process as a constrained maximization problem by limiting the unbounded loss to an $\ell_2$-norm sphere by a designated reference model that may be difficult to obtain. FUKD \cite{wu2022federated} removes the contribution of a target client by subtracting the historical parameter updates and recovering the model performance through knowledge distillation \cite{hinton2015distilling}, necessitating unlabeled data on the server. Additionally, MoDe \cite{zhao2023federated} 
adjusts pre-trained model parameters through two phases—knowledge erasure and memory guidance—to reduce discriminability for target forgotten data and restore performance, dependent on the initial state of the degraded model. These limitations underscore the necessity for a more effective FU framework that addresses these challenges without compromising efficiency or privacy.

To make better use of the information contained in the teacher network, since \cite{romero2014fitnets}, most knowledge distillation
shifted from output distillation to feature distillation \cite{heo2019comprehensive,zhao2022decoupled}, showing superior performance on various tasks. However, existing knowledge distillation-based unlearning methods focus on merely output distillation \cite{zhao2023federated,chundawat2023can,kurmanji2024towards}. MoDe \cite{zhao2023federated} achieves unlearning by using a degraded model (i.e., a ``Bad Teacher") that has never been trained on the forgotten data, to generate pseudo labels for the student model on forgotten data. Chundawat et al. \cite{chundawat2023can} employ a teacher-student objective that minimizes KL-Divergence between the output of the ``Bad Teacher" and the student, encouraging the student model to align closely with the ``Bad Teacher" on the forgotten set. Similarly, SCRUB \cite{kurmanji2024towards} suggests maximizing KL-Divergence to encourage the student model to ``move away" from the trained teacher model on forgotten data. To the best of our knowledge, no method has yet considered encouraging the student model to learn from the ``Bad Teacher" at the feature level to guarantee a higher level forgetting, marking an opportunity for innovation in unlearning.

In this paper, we delve into the paradigm of Federated Client Unlearning and present the first FU framework in medical imaging to ensure the right of a target client to remove the contribution of their data from a trained global model efficiently. We use Model-Contrastive Unlearning (MCU) to encourage the model to perform similarly with a downgraded model and differently with the trained global model on the forgotten data, which pioneers a step towards achieving unlearning at the feature level. Meanwhile, to preserve the generalized knowledge and only remove client-specific knowledge of the target client, we use Frequency-Guided Memory Preservation (FGMP) to preserve the low-frequency components of the trained model, ensuring a rapid post-training on the remaining clients. We validate our proposed method on two real-world tasks, including intracranial hemorrhage (ICH) diagnosis and skin lesion diagnosis. Extensive experiments demonstrate that our method outperforms a number of state-of-the-art FU methods, without compromising privacy, and with an expected speed-up of 10-15 times compared with retraining from scratch.


\begin{figure}[t!]
\centering
\includegraphics[width=0.95\textwidth]{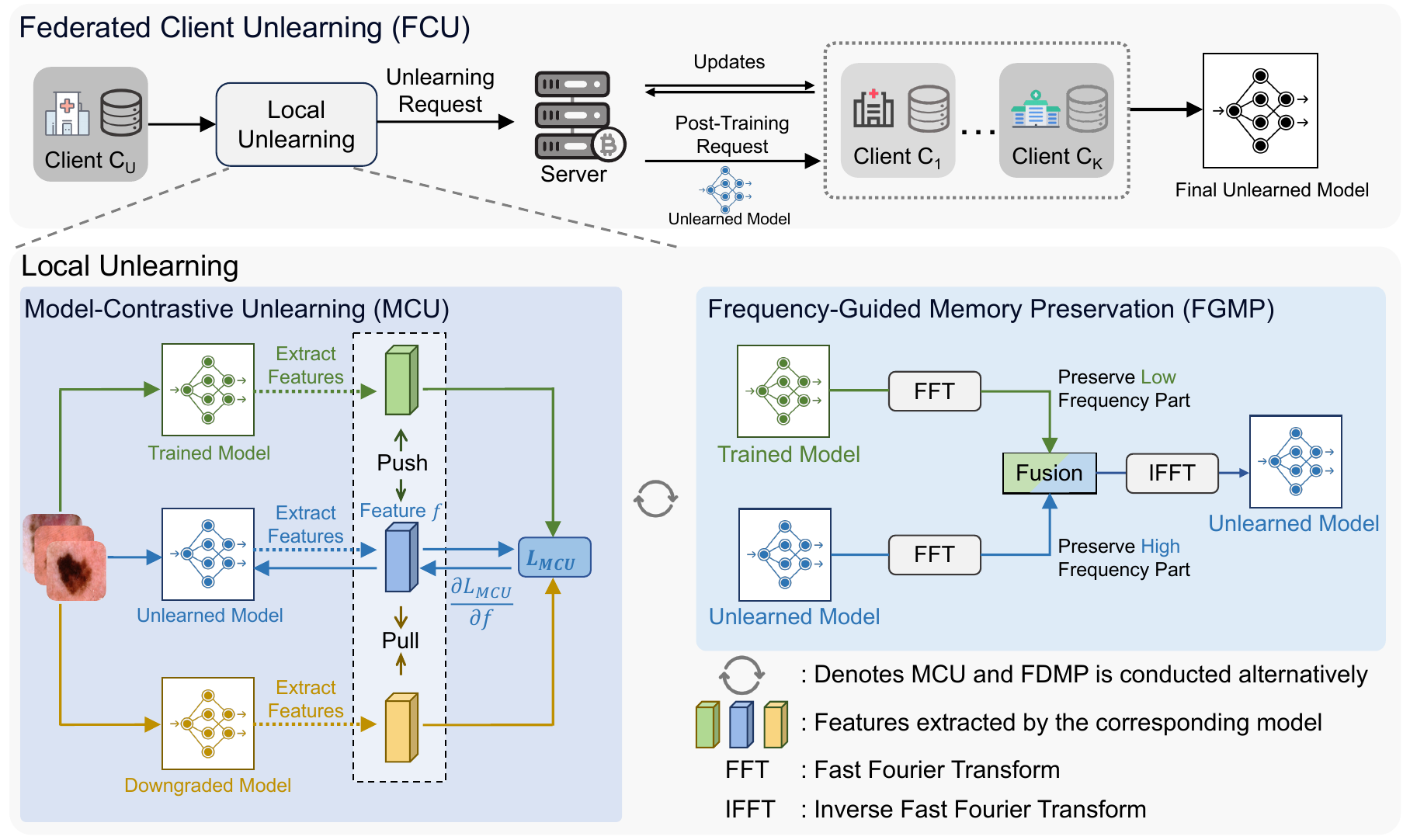}
\caption{Overview of Federated Client Unlearning (FCU): The target client conducts unlearning locally and updates the server with the unlearned model, which then serves as the initial model for post-training on the remaining clients.} \label{fig1}
\end{figure}

\section{Methodology}
\subsection{Preliminaries and Overview}

We adopt a typical Federated Unlearning (FU) scenario as described in prior works \cite{halimi2022federated,liu2021federaser}, involving $K$ clients $\mathbf{C}= \{ C_1,C_2,\ldots, C_K \}$ and a central server participating in FL. The dataset held by each client is denoted as $\mathbf{D}= \{ D_1,D_2,\ldots, D_K \}$. Suppose that after $t$ rounds of FL, each client possesses a trained global model $\mathbf{M}_{tr}$. We refer to the client $C_{u}$ wants to opt out as the \textit{target client}, where $C_{u}$ requests to remove the contribution of their data $D_{u}$ from $\mathbf{M}_{tr}$. Following \cite{liu2021federaser,che2023fast,halimi2022federated}, the goal of this study is to unlearn the $D_{u}$, effectively eliminating its influence from $\mathbf{M}_{tr}$ to produce an unlearned model $\mathbf{M}_{un}$. 

Our proposed framework FCU is presented in Fig.~\ref{fig1}. The target client $C_{u}$ initiates the unlearning process by performing local unlearning to generate the initial unlearned model $\mathbf{M}_{un}$, where the proposed Model-Contrastive Unlearning (MCU) is to make $\mathbf{M}_{un}$ perform similarly with a model that has never trained on $D_{u}$ and differently with the trained model $\mathbf{M}_{tr}$ at the feature level. Furthermore, to preserve the generalized knowledge and only remove client-specific knowledge of the target client, we use Frequency-Guided Memory Preservation (FGMP) to preserve the low-frequency components of the trained model, thereby achieving a rapid post-training to generate the final unlearned model $\Tilde{\mathbf{M}}_{un}$.

\subsection{Model-Contrastive Unlearning}
The intuition of our Model-Contrastive Unlearning (MCU) is to achieve unlearning by encouraging the unlearned model $\mathbf{M}_{un}$ perform similarly at feature level to a model that has never been trained on $D_{u}$, which we refer to as a downgraded model $\mathbf{M}_{down}$. Model-Contrastive Learning was first proposed in \cite{li2021model}, where they aim to decrease the representation drift between the local model and the global model in FL. In contrast, we propose to encourage the unlearned model to output similar features as a downgraded model $\mathbf{M}_{down}$ (pull) and dissimilar features as the trained global model $\mathbf{M}_{tr}$ (push), where we refer to this process as Model-Contrastive Unlearning (MCU). 

We use a model with the same structure as the $\mathbf{M}_{tr}$ but only pretrained on ImageNet\cite{deng2009imagenet} while not being trained on $D_{u}$ as the downgraded model $\mathbf{M}_{down}$, based on two intuitions: 1) $\mathbf{M}_{down}$ also serves as pretrained model for $\mathbf{M}_{tr}$ in FL training before FU, 2) $\mathbf{M}_{down}$ possess the ability to extract low-level features of images \cite{neyshabur2020being}. In contrast, MoDe\cite{zhao2023federated} chose a randomly initialized model as a degraded model to generate pseudo labels for the unlearned model, where this randomly initialized model may be strongly dependent on the initial weights. Similar to \cite{li2021model}, our Model-Contrastive Unlearning loss can be formulated as: 
\begin{equation}
\mathcal{L}_{mcu} = -\log \left( \frac{\exp(\text{sim}(z, z_{\text{down}}) / \tau)}{\exp(\text{sim}(z, z_{\text{down}}) / \tau) + \exp(\text{sim}(z, z_{\text{tr}}) / \tau)} \right),
\end{equation}

\noindent where $\text{sim}(\cdot,\cdot)$ is a cosine similarity function, $\tau$, known as the temperature, $z$ represents the feature vector extracted by the model for a given input $x$, $z_{\text{down}}$ is the feature vector extracted by the downgraded model $\mathbf{M}_{down}$, and $z_{\text{tr}}$ is the feature vector extracted by the trained global model $\mathbf{M}_{tr}$. 



\subsection{Frequency-Guided Memory Preservation}

The goal of the unlearning process is to remove the specific knowledge of the target client $C_u$ without losing the generalized knowledge already learned in a trained global model $\mathbf{M}_{tr}$. Inspired by the findings of \cite{liu2018frequency,chen2021personalized}, which indicates that the low-frequency components of parameters may reflect the basis for global features across all clients while high-frequency components may contain speciﬁc knowledge for an individual client, we introduce 
Frequency-Guided Memory Preservation (FGMP). FGMP aims to preserve the low-frequency components of $\mathbf{M}_{tr}$ and high-frequency components of the unlearned model $\mathbf{M}_{un}$ to acquire a newly unlearned model after MCU. Intuitively, the newly unlearned model $\mathbf{M'}_{un}$ maintains the generalized knowledge inherited from $\mathbf{M}_{tr}$ and has the specific knowledge from  $D_u$ removed in high-frequency components by MCU.

Specifically, we conduct FGMP every $T_{FGMP}$ iterations (e.g. $T_{FGMP}$ is set to 10 in our experiment) while MCU is continuously executed, resulting in an unlearned model $\mathbf{M}_{un}$. We use FFT to convert the parameters of $\mathbf{M}_{un}$ and $\mathbf{M}_{tr}$ into frequency domain, and preserve the low-frequency part of $\mathbf{M}_{tr}$ and high-frequency part of $\mathbf{M}_{un}$. Afterwards, the newly unlearned model $\mathbf{M'}_{un}$ is constructed by inverse FFT (IFFT). We conduct FFT and IFFT of model parameters similarly as \cite{chen2021personalized}. To clarify, for the weights $w$ in a convolutional layer that has $N$ input channels, $H$ output channels and a kernel with size $d_1 \times d_2$, we reshape $w\in\mathbb{R}^{N\times H\times d_{1}\times d_{2}}$ into a 2-D matrix $w'\in\mathbb{R}^{d1N\times d2H}$ to ease the process of FFT and IFFT. Then, we can obtain the amplitude map 
$\mathcal{F}^A$ and phase map $\mathcal{F}^P$ through the Fourier transform $\mathcal{F}=\mathcal{F}^Ae^{j\mathcal{F}^P}$:
\begin{equation}
    \mathcal{F}(w)(m, n) = \sum_{x,y} w'(x, y)e^{-j2\pi\left(\frac{x}{d_1 N}m + \frac{y}{d_2 H}n\right)}, \quad j^2 = -1.
\end{equation}

To extract the low-frequency components, we define a mask matrix $M$ of the same dimensions as $w'$, $M \in \{0, 1\}^{d_1N \times d_2H}$. $M_{ij} = 1$ for the central region, and $M_{ij} = 0$ elsewhere. The central region is a rectangle centered around the middle of $M$, with dimensions $ \lfloor rd_1N \rfloor \times \lfloor rd_2H \rfloor$, where $r$ is the ratio of the low-frequency part to be preserved, and $ \lfloor \cdot \rfloor$ denotes the floor function to ensure integer dimensions. Hence, the newly unlearned model in the frequency domain that preserves the low frequency part of $\mathbf{M}_{tr}$ and high frequency part of $\mathbf{M}_{un}$  can be formulated as: $\hat{\mathcal{F}}^{A}(w'_{adj})) = M \odot \mathcal{F}^{A}(w_{pre}) + (1-M)\odot \mathcal{F}^{A}(w_{adj}),$ where $\odot$ means element-wise multiplication. Finally, we apply IFFT $\mathcal{F}^{-1}$ to convert the amplitude and phase maps back to the parameter as $w'_{adj}=\mathcal{F}^{-1}(\hat{\mathcal{F}}^{A}(w'),\mathcal{F}^{P}(w'))$. 

After FGMP, we obtain a newly unlearned model $\mathbf{M'}_{un}$. Intuitively, the high-frequency components of $\mathbf{M}_{un}$ have been made to forget the specific knowledge of the target client due to the application of MCU, while the low-frequency components retain the generalized knowledge. This selective retention and forgetting forms a solid foundation for post-training, ensuring that the model preserves its ability to generalize well while removing client-specific knowledge.

\subsection{Overall of FCU framework}

After local unlearning, the server sends the unlearned model $\mathbf{M}_{un}$ to the remaining clients request them to conduct post-training using FedAvg \cite{mcmahan2017communication}, where $\mathbf{M}_{un}$ serves as the initial global model and the global objective can be formulated as: $\min\limits_{W} \mathbf{L}(W) = \sum_{k \in \{1, \ldots, K\} \setminus \{u\}} \mathcal{L}_{k}(W)$, where $\mathcal{L}_{k}$ is the local objective of client $C_k$, and $W$ is the parameters of the global model. The global model parameter $W$ is iteratively updated with the aggregation of local models on the remaining clients, which is defined as \(W^{t+1} = \sum_{k \in \{1, \ldots, K\} \setminus \{u\}} \frac{n_k}{n - n_u} {W}_k^{t},\) where: ${W}^{t+1}$ represents the updated global model parameters, $n_k$ and $n_u$ denote the sample sizes of the \(k\)-th client and client \(u\), respectively, $n$ is the total sample size from all clients, and ${W}_k^{t}$ are the parameters from the \(k\)-th client's model. Due to the memory preservation achieved by FGMP, our post-training can efficiently restore model performance on remaining datasets with a few rounds.

\section{Experiments}

\subsection{Experiment Setup}
\noindent \textbf{Datasets.} We evaluated our method on two public real-world medical image classification tasks:1) Intracranial hemorrhage (ICH) diagnosis. We use the RSNA-ICH dataset \cite{flanders2020construction} and follow \cite{jiang2023client} to perform the binary diseased-or-healthy classiﬁcation, and randomly sample 25,000 slices. 2) ISIC2018 skin lesion diagnosis. We conducted skin lesion diagnosis with HAM10000 \cite{tschandl2018ham10000}, which contains 10,015 dermoscopy images. Training, validation and testing sets for both datasets were divided into 7:1:2. For both tasks, to simulate heterogeneous multi-source data, following \cite{shang2022federated}, Dirichlet distribution,i.e. $Dir(\alpha =1.0)$, is used to divide the training set to 5 clients.

\noindent \textbf{Implementation Details.} For both tasks, We used DenseNet121\cite{huang2017densely} as the backbone. The network was optimized by Adam optimizer where the momentum terms were set to 0.9 and 0.99, with learning rate set to $1e^{-5}$ at target client and $1e^{-4}$ for other clients. The total batch size was 64 in both local training and local unlearning, the local unlearning iterations were 100. The temperature $\tau$ in MCU loss was 0.5 by default like \cite{li2021model}. The interval $T_{FGMP}$ to execute FGMP in MCU was set to 10. During post-training,  the local training iterations were 20, and the total communication rounds were 10. The images in both tasks were resized to 224$\times$224. For Task 1, data augmentation included a combination of random flip, rotation, translation, scaling, and gaussian blur. For Task 2, we employed the random flip, rotation, and translation.

 \noindent \textbf{Evaluation Metrics.} \label{metric} We adopt four widely recognized metrics as in recent representative FU study FFMU \cite{che2023fast} and other commonly recognized metrics to assess machine unlearning performance across three dimensions: \textbf{1) Fidelity} that assesses whether the unlearning methods preserve the original model's performance. This includes measuring F1-score, Accuracy, errors on the retained data $Error^r$ (evaluated on $D_r$, the data held by the remaining clients $\mathbf{C} \setminus C_u$), and errors on the test dataset $Error^t$. \textbf{2) Efficacy} that evaluates the success of an FU method in eradicating the influence of the target client's data, $D_u$. This is gauged by $Error^f$, the classification errors on the forgotten dataset $D_f$ (i.e. the dataset on the target client $C_u$). A model's $Error^f$ close to that of a retrained model (which has never encountered $D_f$) is considered favorable \cite{che2023fast,golatkar2020eternal,fu2021knowledge}. \textbf{3) Efficiency} that measures the reduction in communication and computational overheads by quantifying runtime, with each method trained to convergence for a fair comparison. All the results were averaged over 3 runs.


\begin{table}[t!]
\centering
\caption{Comparison with state-of-the-art methods on ICH diagnosis. * denotes the retrained model is regarded as the gold-standard for $Error^f$.}
\label{tab:1}
\begin{tabular}{c|cccc|c|c}
\toprule
\multirow{2}{*}{Methods} & \multicolumn{4}{c|}{Fidelity} & \multicolumn{1}{c|}{Efficacy} & \multicolumn{1}{c}{Efficiency} \\
\cmidrule{2-7}
& Accuracy & F1 & $Error^t$ & $Error^r$ & $Error^f$ & Runtime (s) \\
\midrule
Origin & 86.84 & 86.84 & 13.15 & 8.44 & 6.59${(-5.7)}$  & 3612 \\
Retrain & 85.52 & 85.40 & 14.48 & 10.60 & 12.28*${(0.0)}$  & 2873 \\
\midrule
Finetune & 86.11 & 86.01 & 13.88 & \textbf{7.86} & 7.25${(-5.0)}$  & 301 \\
FFMU \cite{che2023fast} & 85.09 & 84.98 & 14.90 & 10.28 & 9.79${(-2.5)}$ & 249 \\
MoDe \cite{zhao2023federated} & 80.88 & 80.62 & 19.12 & 14.59 & 18.34${(+6.1)}$  & 509 \\
UPGA \cite{halimi2022federated} & 78.96 & 78.93 & 21.04 & 20.43 & 20.25${(+8.0)}$ & 1031 \\
FedEraser \cite{liu2021federaser} & 84.32 & 84.30 & 15.67 & 11.72 & 15.29${(+3.0)}$ & 906 \\
FUKD \cite{wu2022federated} & 83.66 & 83.54 & 16.34 & 12.49 & 15.63${(+3.4)}$ &  1347 \\
\midrule
w/o FGMP & 85.42 & 85.36 & 14.58 & 10.21  & 11.80${(\mathbf{-0.5})}$ & 1831 \\
w/o post-training & 82.57 & 82.53 & 17.42 & 12.06 & 16.80${(+4.5)}$ & \textbf{14} \\
Ours & \textbf{86.40} & \textbf{86.32} & \textbf{13.60} & 8.11 & 11.37${(-0.9)}$  & 177 \\
\bottomrule
\end{tabular}
\end{table}

\subsection{Comparison with state-of-the-arts}

We compared our method with the trained global model denoted as \textbf{origin}, the model \textbf{finetuned} on the trained global model as a baseline, and the model \textbf{retrained} from scratch as gold-standard for efficacy \cite{che2023fast,golatkar2020eternal,fu2021knowledge}. Besides, we compared with five recent state-of-the-art (SOTA) methods, including \textbf{FFMU} \cite{che2023fast} applying random gradient quantization, \textbf{MoDe} \cite{zhao2023federated} utilizing a degraded model to unlearn, \textbf{UPGA} \cite{halimi2022federated} employing projected gradient ascent to maximize the empirical loss on the target client, \textbf{FedEraser} \cite{liu2021federaser} eliminating the inﬂuence of the target client by historical parameter updates iteratively, and \textbf{FUKD} \cite{wu2022federated} that erases the contributions of clients by subtracting the historical parameter updates and restore the performance by knowledge distillation.

The quantitative results for two tasks are presented in Table ~\ref{tab:1} and Table ~\ref{tab:2}. Our method leads in fidelity, achieving improvements of approximately 1.5\% in accuracy and 1.6\% in F1 score for Task 1, and 1.9\% in accuracy and 5.0\% in F1 score for Task 2, compared to the second-best FU method, showing robustness in model performance maintenance. In efficacy, it nearly matches the retrained model considered the standard in forgetting \cite{che2023fast,golatkar2020eternal,fu2021knowledge}, ensuring effective unlearning. Besides, we proved that the finetune method failed to achieve unlearning. For Efficiency, we see an impressive runtime reduction, achieving roughly a 10 to 15 times speed-up compared with retraining from scratch.

\begin{table}[t!]
\centering
\caption{Comparison with state-of-the-art methods on Skin Lesion diagnosis. * denotes the retrained model is regarded as the gold-standard for $Error^f$. }
\label{tab:2}
\begin{tabular}{c|cccc|c|c}
\toprule
\multirow{2}{*}{Methods} & \multicolumn{4}{c|}{Fidelity} & \multicolumn{1}{c|}{Efficacy} & \multicolumn{1}{c}{Efficiency} \\
\cmidrule{2-7}
& Accuracy & F1 & $Error^t$ & $Error^r$ & $Error^f$ & Runtime (s) \\

\midrule
Origin & 79.98 & 57.90 & 20.01 & 9.66 & 29.95${(-5.4)}$ & 2469 \\
Retrain & 81.52 & 54.76 & 18.47 & 5.40 & 35.37*${(0.0)}$ & 2038 \\
\midrule
Finetune & 80.52 & 55.43 & 19.47 & 8.43 & 30.49${(-4.9)}$ & 289 \\
FFMU \cite{che2023fast} & 79.67 & 44.64 & 20.33 & 10.32 & 30.99${(-4.4)}$ & 176 \\
MoDe \cite{zhao2023federated} & 73.04 & 33.63 & 26.96 & 16.55 & 50.84${(+15.5)}$  & 431 \\
UPGA \cite{halimi2022federated} & 75.08 & 38.43 & 24.91 & 17.51 & 48.28${(+12.9)}$ & 899 \\
FedEraser \cite{liu2021federaser} & 80.13 & 53.64 & 19.87 & 8.72 & 37.49${(+2.1)}$ & 672 \\
FUKD \cite{wu2022federated} & 78.11 & 41.27 & 21.89 & 12.69 & 39.72${(+4.4)}$ & 1132 \\
\midrule

w/o FGMP & 80.91 & 53.92 & 19.09 & 6.69 & 33.69${(-1.6)}$ & 1579 \\
w/o post-training & 76.83 & 40.79 & 23.16 & 13.85 & 42.58${(+7.2)}$  & \textbf{13} \\
Ours & \textbf{81.67} & \textbf{56.32} & \textbf{18.33} & \textbf{8.22} & 34.45${(\mathbf{-0.9)}}$ & 156 \\
\bottomrule

\end{tabular}
\end{table}

\begin{figure}[t!]
\centering
\includegraphics[width=1\textwidth]{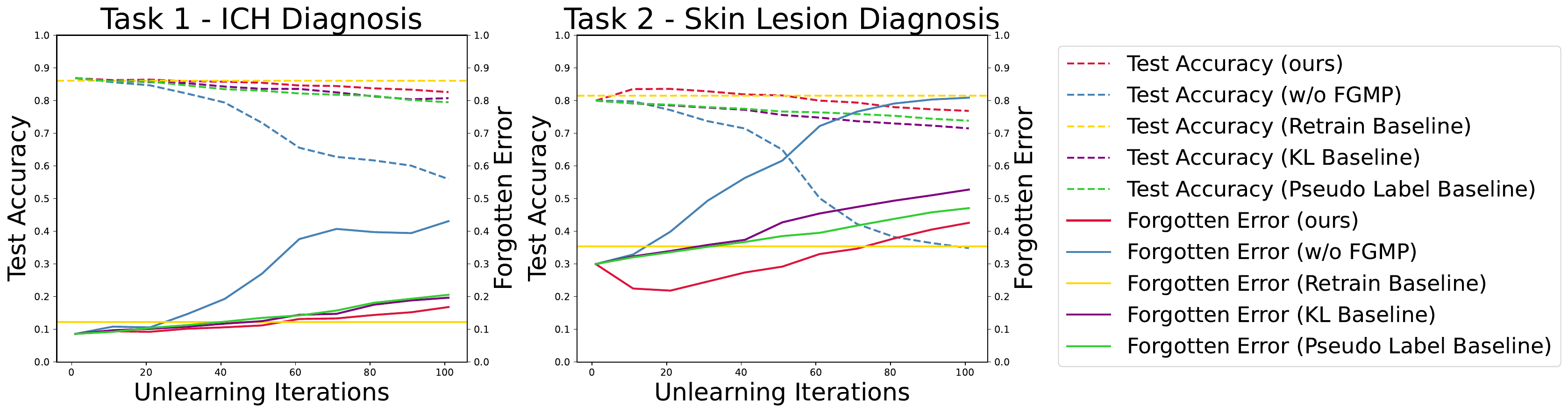}
\caption{Ablation study of FGMP and MCU across local unlearning iterations, showing forgotten $Error^f$ and Accuracy for the two tasks.}
\label{fig2}
\end{figure}

\noindent \textbf{Ablation study.} 
We conducted ablation studies to assess the effectiveness of the primary components of our FCU framework. As shown in Table ~\ref{tab:1} and Table ~\ref{tab:2}, the performance drops significantly without post-training, and the runtime increases a lot without FGMP. The effectiveness of our FGMP in facilitating local unlearning across various iterations is demonstrated in Fig.~\ref{fig2}, which can maintain the performance at its maximum level. Without FGMP, there is a noticeable decline in test accuracy even before the model has fully executed unlearning across the entire dataset. Besides, we replace our MCU with KL-based knowledge distillation \cite{chundawat2023can} and pseudo label-based knowledge distillation \cite{zhao2023federated}, which shows that our MCU can better preserve the performance on test set when achieve at similar forgotten error.

\section{Conclusion}
We present the first federated unlearning framework in medical imaging, which facilitates the right for a client to be forgotten. In the local unlearning phase, our FCU utilizes Model-Contrastive unlearning (MCU) to encourage the model to perform similarly to a model that has never seen the forgotten data at the feature level. To preserve the generalized knowledge, we use Frequency-Guided Memory Preservation (FGMP) to preserve the low-frequency components of the trained global model, ensuring a smooth forgetting process. Benefited from FGMP, our FCU framework quickly restores performance with minimal post-train rounds, achieving a 10-15 times speed-up over retraining from scratch, while demonstrating remarkable federated unlearning effectiveness for medical imaging.

\begin{credits}

\subsubsection{\ackname}
This work was supported by the Hong Kong Innovation and Technology Fund (Project No. MHP/002/22)  and and HKUST 30 for 30 Research Initiative Scheme (No. 3030\_024). 

\subsubsection{\discintname}
The authors have no competing interests to declare that are relevant to the content of this article.
\end{credits}
%
%
%
\bibliographystyle{splncs04}
\bibliography{reference}

\begin{thebibliography}{10}
\providecommand{\url}[1]{\texttt{#1}}
\providecommand{\urlprefix}{URL }
\providecommand{\doi}[1]{https://doi.org/#1}

\bibitem{bourtoule2021machine}
Bourtoule, L., Chandrasekaran, V., Choquette-Choo, C.A., Jia, H., Travers, A., Zhang, B., Lie, D., Papernot, N.: Machine unlearning. In: 2021 IEEE Symposium on Security and Privacy (SP). pp. 141--159. IEEE (2021)

\bibitem{cao2015towards}
Cao, Y., Yang, J.: Towards making systems forget with machine unlearning. In: 2015 IEEE symposium on security and privacy. pp. 463--480. IEEE (2015)

\bibitem{che2023fast}
Che, T., Zhou, Y., Zhang, Z., Lyu, L., Liu, J., Yan, D., Dou, D., Huan, J.: Fast federated machine unlearning with nonlinear functional theory. In: Fortieth International Conference on Machine Learning (2023)

\bibitem{chen2021personalized}
Chen, Z., Zhu, M., Yang, C., Yuan, Y.: Personalized retrogress-resilient framework for real-world medical federated learning. In: Medical Image Computing and Computer Assisted Intervention--MICCAI 2021: 24th International Conference, Strasbourg, France, September 27--October 1, 2021, Proceedings, Part III 24. pp. 347--356. Springer (2021)

\bibitem{chundawat2023can}
Chundawat, V.S., Tarun, A.K., Mandal, M., Kankanhalli, M.: Can bad teaching induce forgetting? unlearning in deep networks using an incompetent teacher. In: Proceedings of the AAAI Conference on Artificial Intelligence. vol.~37, pp. 7210--7217 (2023)

\bibitem{deng2009imagenet}
Deng, J., Dong, W., Socher, R., Li, L.J., Li, K., Fei-Fei, L.: Imagenet: A large-scale hierarchical image database. In: 2009 IEEE conference on computer vision and pattern recognition. pp. 248--255. Ieee (2009)

\bibitem{Deng2023}
Deng, Z., Luo, L., Chen, H.: Scale federated learning for label set mismatch in medical image classification. In: International Conference on Medical Image Computing and Computer-Assisted Intervention. pp. 118--127. Springer (2023)

\bibitem{flanders2020construction}
Flanders, A.E., Prevedello, L.M., Shih, G., Halabi, S.S., Kalpathy-Cramer, J., Ball, R., Mongan, J.T., Stein, A., Kitamura, F.C., Lungren, M.P., et~al.: Construction of a machine learning dataset through collaboration: the rsna 2019 brain ct hemorrhage challenge. Radiology: Artificial Intelligence  \textbf{2}(3),  e190211 (2020)

\bibitem{fu2021knowledge}
Fu, S., He, F., Tao, D.: Knowledge removal in sampling-based bayesian inference. In: International Conference on Learning Representations (2021)

\bibitem{golatkar2020eternal}
Golatkar, A., Achille, A., Soatto, S.: Eternal sunshine of the spotless net: Selective forgetting in deep networks. In: Proceedings of the IEEE/CVF Conference on Computer Vision and Pattern Recognition. pp. 9304--9312 (2020)

\bibitem{halimi2022federated}
Halimi, A., Kadhe, S., Rawat, A., Baracaldo, N.: Federated unlearning: How to efficiently erase a client in fl? arXiv preprint arXiv:2207.05521  (2022)

\bibitem{harding2019understanding}
Harding, E.L., Vanto, J.J., Clark, R., Hannah~Ji, L., Ainsworth, S.C.: Understanding the scope and impact of the california consumer privacy act of 2018. Journal of Data Protection \& Privacy  \textbf{2}(3),  234--253 (2019)

\bibitem{heo2019comprehensive}
Heo, B., Kim, J., Yun, S., Park, H., Kwak, N., Choi, J.Y.: A comprehensive overhaul of feature distillation. In: Proceedings of the IEEE/CVF International Conference on Computer Vision. pp. 1921--1930 (2019)

\bibitem{hinton2015distilling}
Hinton, G., Vinyals, O., Dean, J.: Distilling the knowledge in a neural network. arXiv preprint arXiv:1503.02531  (2015)

\bibitem{huang2017densely}
Huang, G., Liu, Z., Van Der~Maaten, L., Weinberger, K.Q.: Densely connected convolutional networks. In: Proceedings of the IEEE conference on computer vision and pattern recognition. pp. 4700--4708 (2017)

\bibitem{jiang2023client}
Jiang, M., Zhong, Y., Le, A., Li, X., Dou, Q.: Client-level differential privacy via adaptive intermediary in federated medical imaging. In: International Conference on Medical Image Computing and Computer-Assisted Intervention. pp. 500--510. Springer (2023)

\bibitem{kurmanji2024towards}
Kurmanji, M., Triantafillou, P., Hayes, J., Triantafillou, E.: Towards unbounded machine unlearning. Advances in Neural Information Processing Systems  \textbf{36} (2024)

\bibitem{li2021model}
Li, Q., He, B., Song, D.: Model-contrastive federated learning. In: Proceedings of the IEEE/CVF conference on computer vision and pattern recognition. pp. 10713--10722 (2021)

\bibitem{li2021fedbn}
Li, X., Jiang, M., Zhang, X., Kamp, M., Dou, Q.: Fedbn: Federated learning on non-iid features via local batch normalization. arXiv preprint arXiv:2102.07623  (2021)

\bibitem{liu2021federaser}
Liu, G., Ma, X., Yang, Y., Wang, C., Liu, J.: Federaser: Enabling efficient client-level data removal from federated learning models. In: 2021 IEEE/ACM 29th International Symposium on Quality of Service (IWQOS). pp. 1--10. IEEE (2021)

\bibitem{liu2021federated}
Liu, Q., Yang, H., Dou, Q., Heng, P.A.: Federated semi-supervised medical image classification via inter-client relation matching. In: Medical Image Computing and Computer Assisted Intervention--MICCAI 2021: 24th International Conference, Strasbourg, France, September 27--October 1, 2021, Proceedings, Part III 24. pp. 325--335. Springer (2021)

\bibitem{liu2018frequency}
Liu, Z., Xu, J., Peng, X., Xiong, R.: Frequency-domain dynamic pruning for convolutional neural networks. Advances in neural information processing systems  \textbf{31} (2018)

\bibitem{mcmahan2017communication}
McMahan, B., Moore, E., Ramage, D., Hampson, S., y~Arcas, B.A.: Communication-efficient learning of deep networks from decentralized data. In: Artificial intelligence and statistics. pp. 1273--1282. PMLR (2017)

\bibitem{neyshabur2020being}
Neyshabur, B., Sedghi, H., Zhang, C.: What is being transferred in transfer learning? Advances in neural information processing systems  \textbf{33},  512--523 (2020)

\bibitem{romandini2024federated}
Romandini, N., Mora, A., Mazzocca, C., Montanari, R., Bellavista, P.: Federated unlearning: A survey on methods, design guidelines, and evaluation metrics. arXiv preprint arXiv:2401.05146  (2024)

\bibitem{romero2014fitnets}
Romero, A., Ballas, N., Kahou, S.E., Chassang, A., Gatta, C., Bengio, Y.: Fitnets: Hints for thin deep nets. arXiv preprint arXiv:1412.6550  (2014)

\bibitem{shang2022federated}
Shang, X., Lu, Y., Huang, G., Wang, H.: Federated learning on heterogeneous and long-tailed data via classifier re-training with federated features. arXiv preprint arXiv:2204.13399  (2022)

\bibitem{singhal2021federated}
Singhal, K., Sidahmed, H., Garrett, Z., Wu, S., Rush, J., Prakash, S.: Federated reconstruction: Partially local federated learning. Advances in Neural Information Processing Systems  \textbf{34},  11220--11232 (2021)

\bibitem{tschandl2018ham10000}
Tschandl, P., Rosendahl, C., Kittler, H.: The ham10000 dataset, a large collection of multi-source dermatoscopic images of common pigmented skin lesions. Scientific data  \textbf{5}(1), ~1--9 (2018)

\bibitem{voigt2017eu}
Voigt, P., Von~dem Bussche, A.: The eu general data protection regulation (gdpr). A Practical Guide, 1st Ed., Cham: Springer International Publishing  \textbf{10}(3152676),  10--5555 (2017)

\bibitem{wu2022federated}
Wu, C., Zhu, S., Mitra, P.: Federated unlearning with knowledge distillation. arXiv preprint arXiv:2201.09441  (2022)

\bibitem{xu2023machine}
Xu, H., Zhu, T., Zhang, L., Zhou, W., Yu, P.S.: Machine unlearning: A survey. ACM Computing Surveys  \textbf{56}(1),  1--36 (2023)

\bibitem{zhao2022decoupled}
Zhao, B., Cui, Q., Song, R., Qiu, Y., Liang, J.: Decoupled knowledge distillation. In: Proceedings of the IEEE/CVF Conference on computer vision and pattern recognition. pp. 11953--11962 (2022)

\bibitem{zhao2023federated}
Zhao, Y., Wang, P., Qi, H., Huang, J., Wei, Z., Zhang, Q.: Federated unlearning with momentum degradation. IEEE Internet of Things Journal  (2023)

\end{thebibliography}

\end{document}